# A geometric sequence of neutral vector boson and charged lepton masses

Bernard F. Riley

NNC Ltd., Birchwood Park, Warrington WA3 6BZ, United Kingdom.
bernard.riley@nnc.co.uk

The neutral vector meson resonances, $\phi$, J/$\psi$, $B_s^*$ and $\Upsilon$, and the tau lepton occupy consecutive mass levels in a geometric series which also includes the muon and the neutral vector boson, Z, in a symmetric arrangement. The mass of the tau lepton is equal to the geometric mean of the $\phi$ and J/$\psi$ masses.

A geometric sequence of neutral vector boson and charged lepton masses has been identified (Figure 1). In this paper, an attempt is made to put that discovery into the context of the model from which it arose.

The quark mass evaluations of the Particle Data Group [1] have been shown [2] to be consistent with the values of 'mass levels' which descend from the Planck Mass (1.2210 ± 0.0009 x10$^{19}$ GeV), $m_P$, in a geometric series of common ratio 2/$\pi$. The mass of the nth level in the spectrum is defined as

$$m_n = \left(\frac{\pi}{2}\right)^{-n} m_P \qquad (1)$$

Also in [2], hadrons were shown to occupy 'higher order' mass levels, of fractional (half-integer, quarter-integer, eighth-integer, etc) n. Doublet states were found to lie either side of mass levels. 'Mass partnerships' occur between particles of, normally, equal strangeness but different I, $I_3$ or J. The mass difference characteristic of the partnership is approximately equal to the mass of a principal (integer n) or low order level. In several partnerships involving strange hadrons ($\Lambda$-$\Sigma^0$, $\Sigma^+$-$\Sigma^-$, $\Sigma^0$-$\Sigma(1385)^0$, K*-$\Sigma^0$ and $\phi$-$\Xi$), the mass difference is equal, within the small uncertainty, to the mass of a principal level. The muon and tau lepton were proposed as the mass partners of the $\pi$ pseudoscalar meson doublet and the $D_s^\pm$ pseudoscalar meson, respectively. The electron was proposed as the mass partner of the K pseudoscalar meson doublet but displaced in mass from within the main sequence of particles, which extends in mass from $m_\mu$ to $m_Z$.

The mass of the electron is related to the mass of the K* vector meson doublet interstitial level (893.80 MeV), and is given by

$$m_e = \frac{1}{2}\left(\frac{\pi}{2}\right)^{-15} m_{K*} \qquad (2)$$

which has the value 0.51100 ± 0.00036 MeV. The measured value is 0.510998902 ± 0.000000021 MeV [3].

A second, bottom-up, geometric mass series is now constructed, ascending from the muon mass with common ratio $\pi/2$. The mass of the n'th level in the spectrum is defined as

$$m_{n'} = \left(\frac{\pi}{2}\right)^{n'} m_\mu \qquad (3)$$

The main sequence of particles spans 15 principal (integer n') level intervals, as shown in Figure 1. The neutral vector meson resonances with hidden flavour, $\phi$ ($s\bar{s}$), J/$\psi$ ($c\bar{c}$), $B_s^*$ ($s\bar{b}$) and $\Upsilon$ ($b\bar{b}$), the neutral vector boson, Z, and the tau lepton occupy principal and higher order n' levels in an arresting pattern. Values of n' for the vector boson masses [3] do not coincide exactly with n' levels because these particles are constrained to occupy the mass levels described by (1) [2]. The charged $D_s^{*\pm}$ vector meson resonance does not occupy a low order level in the n' sequence. The tau lepton, though, occupies the level with n'=6¼.

As a consequence of their level occupation, the muon and the tau lepton are related in mass by

$$\frac{m_\tau}{m_\mu} = \left(\frac{\pi}{2}\right)^{\frac{25}{4}} \qquad (4)$$

Using Particle Data Group mass values [3], the left hand side of (4) is equal to 16.818 ± 0.003; the right hand side of (4) has the value 16.817. The tau lepton, of mass $1776.99^{+0.29}_{-0.26}$ MeV [3], occupies a level mid-way between those of $\phi$ and J/$\psi$. Its mass is equal to the geometric mean of those two mesons, $G = \sqrt{m_\phi m_{J/\psi}}$, which has the value 1776.83 ± 0.06 MeV.

The views expressed in this paper are those of the author and not necessarily those of NNC Ltd.

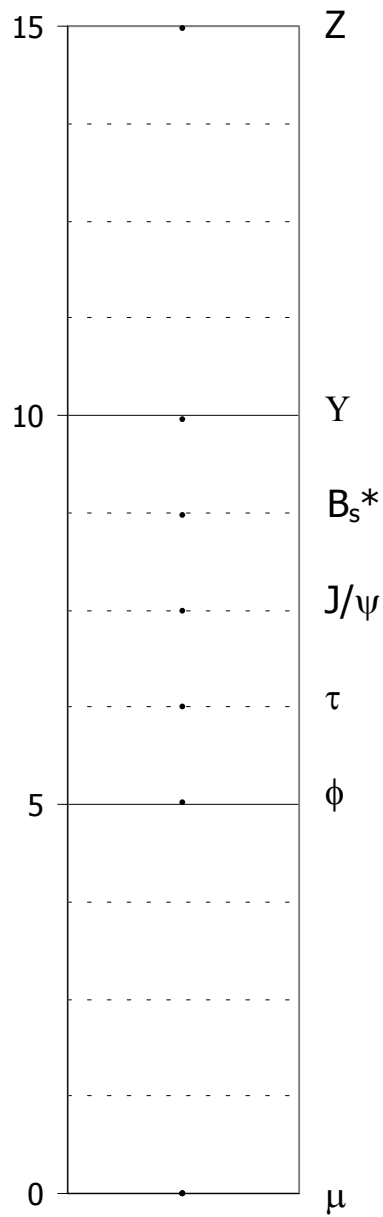

Figure 1: Values of n' in (3), corresponding to the masses of neutral vector meson resonances of hidden flavour, the vector boson, Z, the muon and the tau lepton.